\documentclass[acmsmall,screen,nonacm]{acmart}

\usepackage{booktabs}
\usepackage{tabularx}
\usepackage{array}

\graphicspath{{./}}
\newcolumntype{Y}{>{\raggedright\arraybackslash}X}
\newcommand{\pathcounts}[3]{L#1/\allowbreak{}F#2/\allowbreak{}P#3}
\setcopyright{none}
\renewcommand\footnotetextcopyrightpermission[1]{}

\title[NEURON-Fabric]{NEURON-Fabric: Architecture-Runtime Co-Design for Controlled Low-Bit Gradient Communication}

\author{Ziqiang Wang}
\email{ziqiangwang@cmail.carleton.ca}
\affiliation{%
  \department{Department of Systems and Computer Engineering}
  \institution{Carleton University}
  \city{Ottawa}
  \state{Ontario}
  \country{Canada}
}

\author{Changcheng Huang}
\email{ChangChengHuang@cunet.carleton.ca}
\affiliation{%
  \department{Department of Systems and Computer Engineering}
  \institution{Carleton University}
  \city{Ottawa}
  \state{Ontario}
  \country{Canada}
}

\author{Chung-Horng Lung}
\email{ChungLung@cunet.carleton.ca}
\affiliation{%
  \department{Department of Systems and Computer Engineering}
  \institution{Carleton University}
  \city{Ottawa}
  \state{Ontario}
  \country{Canada}
}

\begin{document}

\begin{abstract}
Large-scale neural-network training repeatedly aggregates gradients across devices, making communication a central cost in distributed learning. Low-bit gradient aggregation is an attractive way to reduce this cost, but applying it as a static replacement for full-precision communication can destabilize training because the safe precision choice depends on training phase, model structure, runtime bucketization, and the communication substrate.

This article presents NEURON-Fabric, a profile-guided runtime system for controlled low-bit gradient aggregation. Instead of treating compression as a fixed tensor transformation, NEURON-Fabric decides at runtime when low-bit aggregation should be admitted, when it should fall back to FP32, and which model regions are eligible for each route. The system combines calibrated operating profiles, model-aware runtime bindings, online training-health monitoring, and reducer-capacity checks so that low-bit communication is applied only when it is both numerically safe and architecturally useful.

Across vision, Transformer, and autoregressive language-model workloads, NEURON-Fabric validates the runtime path from calibration to distributed communication-hook execution. Static low-bit communication can collapse training accuracy, while profile-guided control preserves accuracy near full-precision references or calibrated targets and reduces modeled gradient-communication traffic in the evaluated settings. Transformer and billion-parameter language-model checks further show that the same routing and fallback mechanisms execute across model families and multi-node deployments. Reducer-side replay and reducer-path measurements identify when compact sign-count aggregation is expected to reduce communication cost and when endpoint capacity should trigger fallback.

Together, these results position low-bit gradient aggregation as an architecture-runtime co-design problem rather than a standalone compression operator. NEURON-Fabric shows that low-bit communication can be represented as a calibrated runtime policy, mapped to model semantics, monitored during training, and admitted only when the communication substrate can benefit from it. This provides a practical foundation for future production reducers and broader workload-portability studies.
\end{abstract}

\maketitle

\section{Introduction}

Scaling neural-network training across many accelerators turns collective communication into a recurring systems cost. Data-parallel, tensor/model-parallel, and hybrid-parallel training regimes all use collective reductions or closely related collectives to keep replicated, partitioned, or staged training state consistent. This paper focuses on the DistributedDataParallel (DDP) style gradient aggregation path: after local backward computation produces gradients, workers exchange and reduce those gradients before applying a synchronized update. The conventional 32-bit floating-point (FP32) all-reduce path is robust, but it pays full-precision communication cost for every gradient tensor, including model regions that may tolerate lower-bit communication. We use \emph{gradient-communication cost} to mean the time or traffic spent moving and reducing gradients rather than computing them. Low-bit gradient aggregation can reduce this cost, especially on multi-GPU or multi-node systems where network movement is visible.

The difficulty is that safe low-bit execution is not a property of a single compression operator. It depends on training phase, model region, bucket construction, live training health, and the communication substrate. These dependencies arise at different layers of the system. Training phase and live health determine when low-bit communication may be admitted or revoked. Model region determines which parameters are eligible. Bucket construction determines whether those eligible parameters remain separable inside the communication runtime. Finally, the communication substrate determines whether an admitted low-bit route is actually beneficial.

The last two dependencies create the main runtime integration challenge. In PyTorch DDP, bucket construction is a communication optimization rather than a semantic partition of the model: each bucket groups gradient tensors so they can be all-reduced efficiently, and a single bucket may contain both low-bit-eligible tensors and FP32-sensitive tensors such as embeddings, layer norms, biases, or classifier heads. Below the DDP hook, the communication substrate imposes a separate constraint. Even when a tensor is numerically admissible and correctly routed inside a bucket, the hierarchical reducer--the component that combines per-worker gradient information--can lose its advantage if the local aggregation endpoint becomes the bottleneck.

This paper asks the resulting systems question: how should a distributed runtime admit low-bit gradient aggregation only when the workload, model structure, online health signal, and reducer capacity all support it? The earlier NEURON-Fabric conference version introduced the basic low-bit aggregation setting and the Commander/Supervisor control loop for admission and recovery \cite{neuronfabric2026conference}. This journal article extends that concept into an architecture-runtime system: it specifies the profile-to-runtime interface, executes the profile inside real DDP buckets, adds reducer-capacity admission, and validates the bounded systems claim across workloads and cluster deployments.

NEURON-Fabric has five cooperating components. The first is the operating profile, a compact prior configuration map exported by offline calibration. It records admission timing, selected low-bit mode, initial cumulative-sum (CUSUM) change-detector settings, cooldown and readmission guards, a planned bound on low-bit route exposure, and provenance. The profile supplies calibrated prior knowledge, but it does not freeze every online decision.

The second component is Commander admission. Commander starts from the FP32 warm-up window and uses diagnostics collected during the current run to choose whether the candidate communication mode should be admitted for the current workload and model scope. The third component is Supervisor recovery. Supervisor monitors live training health after admission and decides when the admitted low-bit route must fall back to FP32, remain in cooldown, or be readmitted.

The fourth component is runtime binding under real DDP bucketization. A workload/model binding maps parameters and buckets to roles such as backbone, head, embedding, layer norm, bias, low-bit eligible tensor, and FP32-sensitive tensor. The PyTorch DDP communication hook combines this binding with Commander and Supervisor state to route eligible gradients through low-bit aggregation or FP32 fallback. The key mechanism is mixed-bucket per-parameter routing: when one bucket contains both eligible encoder weights and FP32-sensitive tensors, NEURON-Fabric preserves model semantics below bucket granularity instead of quantizing or rejecting the whole bucket.

The fifth component is reducer-aware admission. A low-bit route that is numerically admissible is not automatically profitable for the communication substrate. Payload replay, endpoint-capacity sweeps, contention replay, and executable hierarchical sign-count checks define when a hierarchical reducer can lower gradient-communication cost and when endpoint service becomes the bottleneck. The endpoint-capacity guard is therefore part of the runtime decision: endpoint-bound low-bit routes fall back to FP32.

The evaluation tests NEURON-Fabric as an executable controlled runtime rather than as a static low-bit replacement for FP32 communication. The experiments cover vision workloads, Stanford Sentiment Treebank 2 (SST-2) Transformer workloads~\cite{socher2013sentiment,wang2019glue}, Pythia-based autoregressive language-model scale-out checks~\cite{biderman2023pythia} with matched short-horizon FP32 loss/perplexity anchors, real DDP bucket bindings, topology and training-duration changes, and reducer-path measurements. Appendix Tables~\ref{tab:app-online-controller-details}, \ref{tab:app-runtime-binding}, \ref{tab:app-reducer-support}, \ref{tab:app-eval-environment}, and \ref{tab:app-transformer-support} summarize the supporting evidence and measurement scope.

This paper makes four systems contributions:

\begin{enumerate}
\item \textbf{A controlled low-bit training runtime.} NEURON-Fabric turns low-bit gradient aggregation from a static compression choice into a runtime-controlled execution mode with admission, fallback, cooldown, and readmission.
\item \textbf{A portable profile-to-runtime interface.} The system separates calibrated operating profiles from workload/model bindings, so the same runtime machinery can execute different model families while preserving model semantics inside mixed DDP buckets.
\item \textbf{Reducer-aware admission for distributed deployments.} The runtime connects training-health control to hierarchical sign-count reducer measurements, admitting low-bit only when endpoint capacity and communication conditions make the route useful.
\item \textbf{Scale-out runtime validation.} The framework is evaluated as an executable DDP/NCCL runtime on multi-node A100 deployments, including vision, Transformer, and autoregressive language-model runs; the Pythia checks add matched FP32 loss/perplexity anchors while route-correctness checks and modeled gradient-communication traffic are recorded under real bucketization.
\end{enumerate}

\section{Background and Systems Constraints}

This section gives the technical background needed for the rest of the paper and states the systems constraints that follow from it. Implementation objects are introduced in Section~3; here we establish why controlled low-bit aggregation is a systems problem rather than only a compression rule.

\subsection{Distributed Gradient Communication}

Synchronous distributed training often requires workers to combine tensors so that different devices apply a consistent update or exchange consistent intermediate state. The common collective primitive for this aggregation step is \emph{all-reduce}: every participant contributes a tensor, the system reduces those tensors, typically by summation or averaging, and the reduced tensor is returned to every participant. All-reduce appears in data parallelism, and related collectives such as reduce-scatter and all-gather are also used when optimizer state, parameters, activations, or model state are sharded. This paper evaluates the DDP-style data-parallel case, where FP32 gradient all-reduce remains a basic communication path \cite{bennun2019demystifying,li2020pytorchddp}.

Modern runtimes do not usually all-reduce each parameter tensor separately. PyTorch DDP groups gradients into \emph{communication buckets} to reduce launch overhead and overlap communication with backward computation \cite{li2020pytorchddp}. A bucket is therefore a runtime communication unit. It is formed for scheduling and efficiency, not because all tensors in the bucket play the same role in the model. Any precision-changing communication path must therefore execute inside the communication structure provided by the framework while preserving model semantics below or across bucket boundaries.

\subsection{Low-Bit Gradient Aggregation}

Low-bit gradient aggregation reduces communication by exchanging compact gradient representations instead of full FP32 values. Prior work has explored one-bit, ternary, quantized, and sign-based communication-efficient training \cite{seide2014onebit,alistarh2017qsgd,wen2017terngrad,karimireddy2019error}. These methods differ in encoding, error compensation, and convergence assumptions, but they share a systems motivation: if the communicated representation is much smaller than FP32 and does not harm optimization too much, the gradient-communication cost can decrease.

NEURON-Fabric operates at this communication boundary. It does not change the model architecture, local forward computation, backward gradient computation, or optimizer update rule. Instead, it changes how eligible gradients are aggregated after they are produced and before synchronized replicas apply the update. For the low-bit routes evaluated in this paper, each worker materializes a compact direction payload, such as a gradient sign or ternary direction, rather than communicating the full FP32 gradient value. The runtime or reducer path aggregates these compact directions and returns an aggregate direction to the optimizer-visible gradient buffer. The payload can be much smaller than FP32, but the representation discards magnitude information and can change the optimization trajectory.

To avoid conflating communication primitives with model quantization, and following the earlier NEURON-Fabric conference version~\cite{neuronfabric2026conference}, this paper names the two evaluated low-bit aggregation formats \emph{G-Binary} and \emph{G-Ternary}. G-Binary is a one-bit sign-count route, while G-Ternary adds a zero-gating rule that can suppress selected entries before aggregation. These names do not mean that the model itself is quantized, as in binary-network or low-bit training methods \cite{courbariaux2015binaryconnect,rastegari2016xnor,wen2017terngrad}. They identify two aggressive gradient-payload formats that borrow arithmetic ideas from low-bit learning but apply them only to gradient communication. More conservative INT8, FP8, block-floating-point, or mixed-precision payloads could offer different accuracy/traffic tradeoffs; the contribution here is to treat precision as a policy-selectable runtime dimension rather than a fixed model format.

\subsection{Bucket Semantics and Temporal Safety}

We use \emph{model-semantic unit} to denote a parameter group defined by its role in the model rather than by the communication bucket that happens to contain it. Examples include a convolutional backbone, Transformer encoder block, embedding layer, normalization layer, bias term, or classifier head. Because these groups contribute differently to optimization and representation quality, they can differ in their tolerance to low-bit communication. DDP buckets do not necessarily align with them: a single bucket may contain both low-bit-tolerant tensors and FP32-sensitive tensors. Treating the whole bucket as one precision unit either risks corrupting sensitive tensors or discards useful low-bit opportunity through whole-bucket fallback. A practical runtime therefore needs a way to preserve model semantics within the bucketized communication path.

Precision eligibility is also temporal. Early training can be more fragile than later training; a route that is safe after warm-up may be unsafe before the model reaches a stable region; and a route that is useful during a healthy period may need to be suspended when loss or validation behavior degrades. FP32 is therefore not only a baseline for comparison. In a controlled runtime it is also the warm-up path, recovery path, and semantic reference path. Low-bit admission and recovery must be stateful rather than encoded as static precision labels.

\subsection{Profile and Binding Portability}

Low-bit behavior is workload- and model-dependent. A portable runtime therefore needs an explicit boundary between control knowledge, such as when low-bit communication should be attempted and how conservative recovery should be, and structural model knowledge, such as which parameters belong to the backbone, head, embeddings, normalizations, biases, or other sensitive regions. Portability does not mean a universal zero-shot rule; it means that workload-specific control assumptions and model-specific structural assumptions are explicit, auditable, and replaceable.

\subsection{Reducer Capacity and Endpoint Bottlenecks}

Low-bit aggregation also depends on the communication substrate. Reducing network bytes helps only if the system does not introduce a new bottleneck elsewhere. Hierarchical and in-network aggregation systems show that reduction can be pushed into switches, endpoints, or aggregation services, but these mechanisms are constrained by local bandwidth, processing rate, queueing, and fan-in \cite{graham2016sharp,sapio2021switchml,gebara2021panama}. For a node-local reducer, \emph{fan-in} is the number of workers that submit updates to the same aggregation endpoint during a reducer step. A route is \emph{endpoint-bound} when endpoint service becomes the limiting term: counting compact updates, returning aggregate state, or queueing under high fan-in can take longer than the network transfer saved by the smaller payload. Thus, a numerically admissible low-bit route is not automatically architecturally useful.

\subsection{Problem Statement}

These background facts lead to the central systems problem:

\begin{quote}
Given a distributed-training run and a candidate low-bit aggregation route, decide at runtime when the route should be admitted, when it should be revoked, and when FP32 fallback should be used, while respecting model structure, training health, and reducer capacity.
\end{quote}

The next section maps these constraints onto the NEURON-Fabric architecture: Commander admission, Supervisor recovery, operating profiles, workload/model bindings, mixed-bucket routing, and endpoint-capacity guards.

\section{System Overview}

NEURON-Fabric separates calibration from distributed execution: calibration decides which low-bit route is admissible for a workload/model pair and provides recovery priors; the runtime maps that decision onto model structure, DDP buckets, fallback paths, and reducer-capacity constraints. This section introduces the runtime control loop, the calibration-to-runtime interface, and the training-step path that consumes that interface.

\subsection{Runtime Control Loop}

\begin{figure}[!t]
\centering
\includegraphics[width=\linewidth]{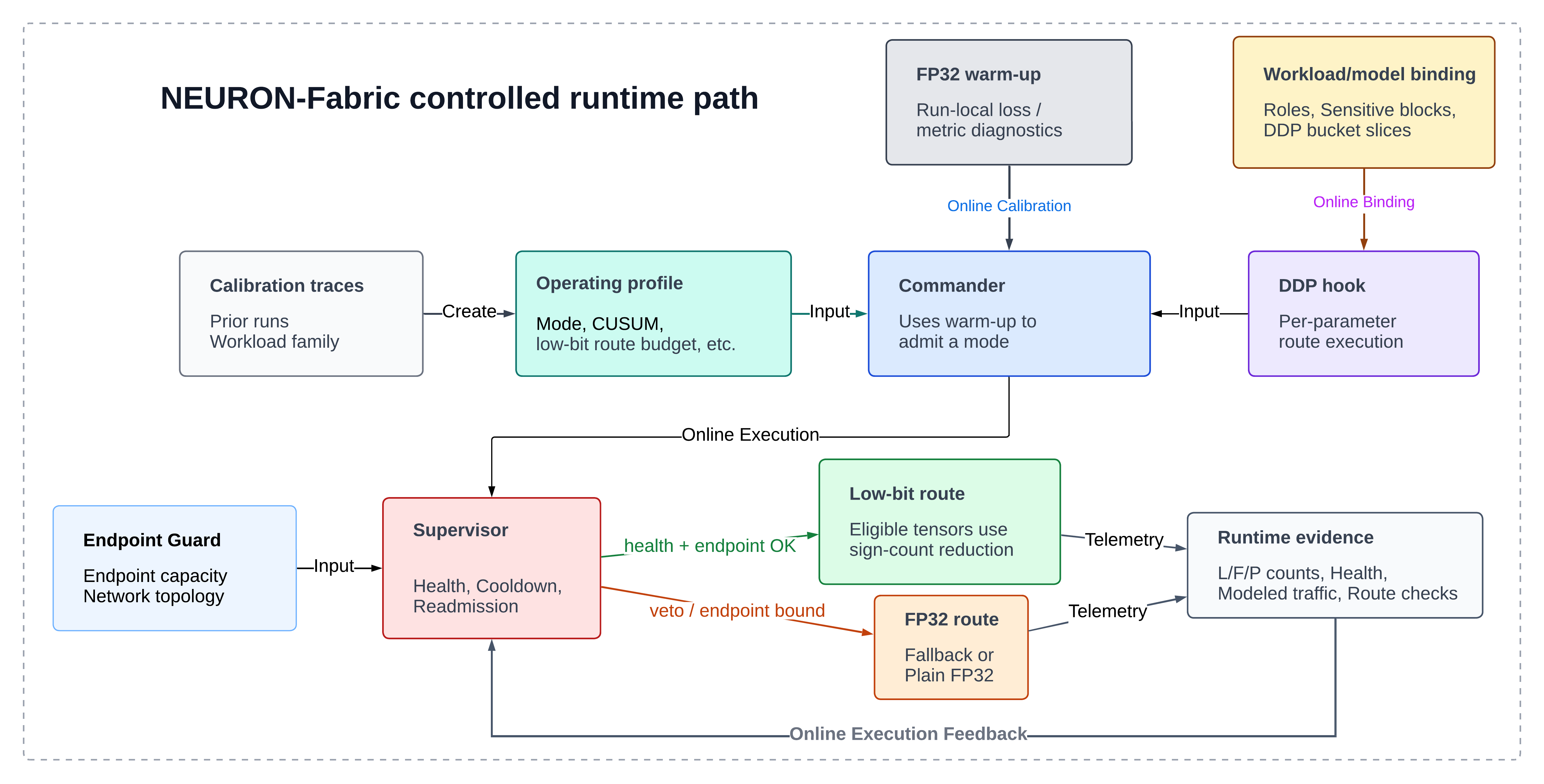}
\caption{Runtime path for controlled low-bit gradient aggregation.
Calibration provides an operating profile, FP32 warm-up provides current-run
diagnostics for Commander admission, the workload/model binding maps the
admitted mode onto DDP bucket contents, and the DDP hook chooses low-bit
reduction or FP32 fallback under Supervisor and endpoint-capacity guards.}
\Description{Block diagram of the NEURON-Fabric runtime path connecting calibration traces, operating profile, FP32 warm-up, Commander admission, workload/model binding, DDP hook routing, Supervisor and endpoint guards, low-bit reduction, FP32 fallback, and runtime evidence logging.}
\label{fig:system-loop}
\end{figure}

Figure~\ref{fig:system-loop} summarizes how calibration, runtime control, model binding, and reducer admission interact during one training run. The remainder of this subsection explains the five components in that loop.

The NEURON-Fabric runtime control loop in Figure~\ref{fig:system-loop} connects five components. Offline calibration exports an operating profile as a calibrated control prior for the run. Commander admission uses FP32 warm-up and the profile to decide whether a candidate low-bit route may be admitted. Supervisor recovery monitors live training health after admission and controls fallback, cooldown, and readmission.

Before per-bucket routing, the distributed runtime loads the workload/model binding, which maps model parameters and DDP buckets to model-semantic precision roles; admitted routes are applied only to the eligible portions. Reducer admission checks whether the communication substrate is expected to benefit from low-bit reduction rather than becoming endpoint-bound. The hook records the selected routes and validation evidence during execution. The following subsections make this loop concrete in two complementary ways: the calibration-to-runtime interface names the objects and signals that cross from calibration into execution, and the runtime-execution walkthrough shows how those inputs are used during one training step.

\subsection{Calibration-to-Runtime Interface}
\label{sec:calibration-runtime-interface}

\begin{figure}[!t]
\centering
\includegraphics[width=\linewidth]{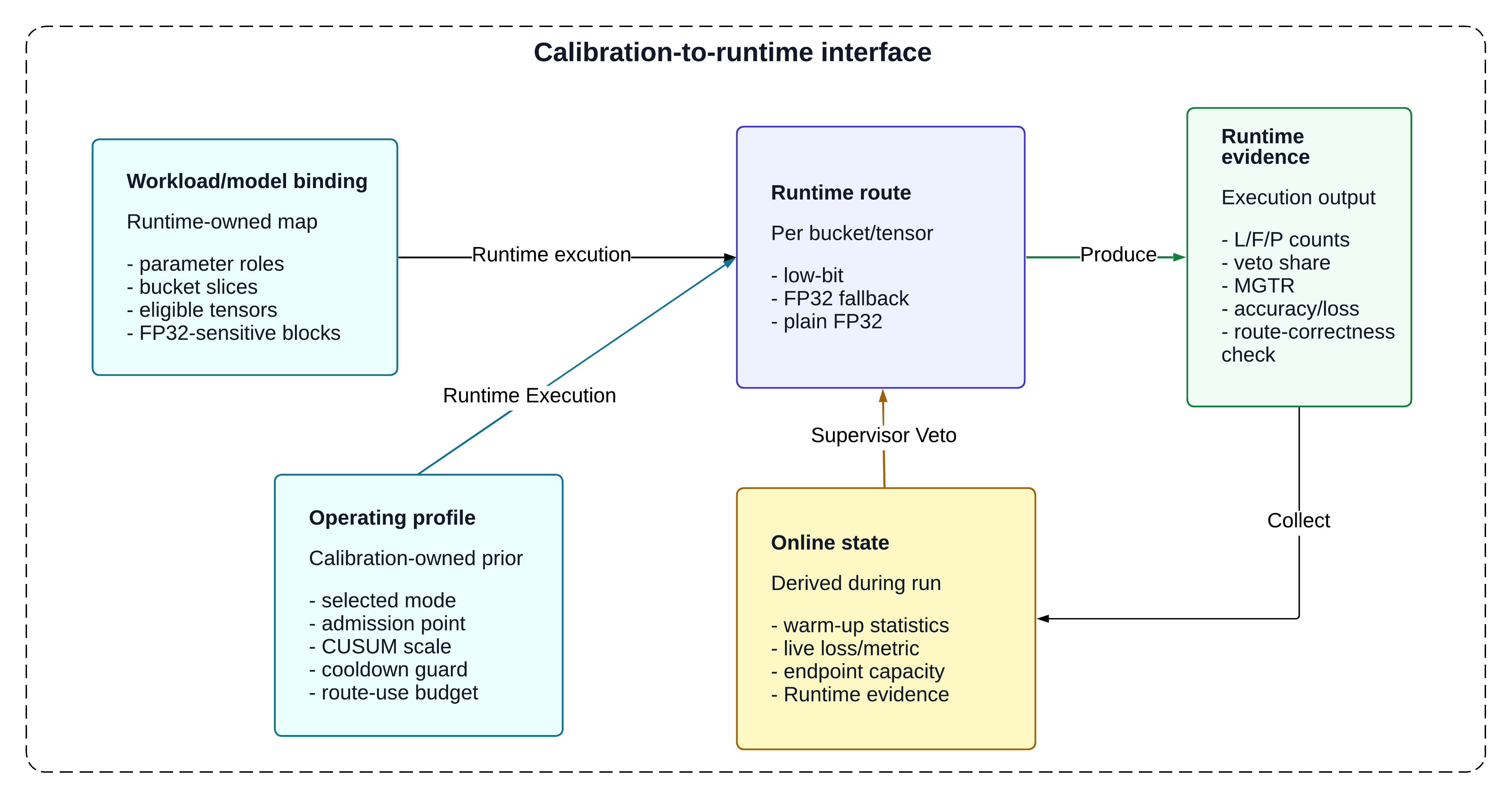}
\caption{Calibration-to-runtime interface. The operating profile and workload/model binding enter execution as fixed inputs, online state is updated during the run, runtime routes are selected per bucket or tensor, and runtime evidence records the observed execution outcome.}
\Description{Diagram of the calibration-to-runtime interface connecting the operating profile and workload model binding to online state, runtime route decisions, and runtime evidence.}
\label{fig:calibration-runtime-interface}
\end{figure}

The control loop requires an explicit boundary between information that is calibrated before a run, state that is observed during a run, decisions made by the communication hook, and evidence recorded after execution. Without this separation, a calibrated profile could be mistaken for an online decision rule, or a runtime log could be treated as prior knowledge. NEURON-Fabric therefore exposes a narrow calibration-to-runtime interface: fixed inputs enter execution, online state changes during training, runtime routes are selected per bucket or tensor, and execution evidence remains an output for auditing, evaluation, and later calibration. Figure~\ref{fig:calibration-runtime-interface} shows this interface, and Table~\ref{tab:calibration-runtime-interface} classifies each item by source, lifetime, and role in route selection.

\begin{table}[t]
\caption{Interface items and runtime status. The profile and binding are inputs to a run; warm-up statistics, live signals, runtime routes, and evidence are produced or updated during execution.}
\label{tab:calibration-runtime-interface}
\scriptsize
\setlength{\tabcolsep}{3pt}
\begin{tabularx}{\linewidth}{YYYY}
\toprule
Interface item & Source & Status during a run & Runtime role \\
\midrule
Workload/model binding & Model architecture plus binding code & Loaded at initialization after preflight & Maps parameters and buckets to backbone, head, sensitive blocks, and low-bit eligibility \\
Operating profile & Calibration traces and profile schema & Loaded at initialization; read-only during a run; swappable between runs & Selects admission point, low-bit mode, initial CUSUM settings, cooldown, guard, exposure bound, and provenance \\
Warm-up health statistics & FP32 training window & Observed online before admission & Derives effective Supervisor scale from current-run loss or metric behavior \\
Live loss or metric signal & Training loop & Dynamic during training & Feeds fallback, cooldown, and readmission decisions \\
Endpoint-capacity guard input & Reducer-path measurements, topology, and endpoint service assumptions & Loaded or updated before route admission; refreshed when hardware, topology, or reducer implementation changes & Decides whether an otherwise admitted low-bit route is expected to reduce communication cost or should fall back to FP32 \\
Runtime route & PyTorch DDP communication hook plus Supervisor state & Dynamic per step and bucket/tensor & Chooses low-bit, FP32 fallback, or plain FP32 route \\
Runtime evidence & Hook instrumentation and validation loop & Experimental output & Records path counts, veto share, modeled gradient-communication traffic, route-correctness checks, accuracy, and validation status \\
\bottomrule
\end{tabularx}
\end{table}

Several evaluation terms in later sections refer to the runtime evidence row in Table~\ref{tab:calibration-runtime-interface}. \emph{Route counts} or \emph{path counts} count hook events by executed path: low-bit (\texttt{L}), FP32 fallback after an admitted low-bit route is vetoed (\texttt{F}), and plain FP32 (\texttt{P}). \emph{Low-bit route share} is the fraction of recorded route events that actually execute the low-bit path, for example \texttt{L/(L+F+P)} when path counts are reported. \emph{Veto share} is the fraction of otherwise eligible route events that the Supervisor sends to FP32 fallback.

\emph{Modeled gradient-communication traffic} is reported as a ratio to the matched FP32 route. It estimates gradient payload traffic under the observed runtime route log and is not a measured wall-clock speedup. For a recorded sequence of route events $E$, NEURON-Fabric reports the modeled gradient-communication traffic ratio (MGTR) as
\[
\mathrm{MGTR}
=
\frac{\sum_{e \in E} \mathrm{bytes}(\mathrm{route}_e)}
     {\sum_{e \in E} \mathrm{bytes}(\mathrm{FP32}_e)} .
\]
The numerator counts the route actually selected by the runtime: FP32 fallback events and FP32-sensitive tensors count as FP32 bytes, while admitted low-bit tensor portions count according to the selected low-bit representation. The denominator replays the same route-event sequence as if every gradient tensor had used FP32 communication. This ratio includes only modeled gradient payload traffic under the observed route log. It excludes dataloader time, forward/backward compute, optimizer work, kernel launch overhead, synchronization delay, collective implementation effects, non-gradient communication, and any end-to-end training throughput measurement. A \emph{route-correctness violation} occurs when the observed route disagrees with the binding and Supervisor state, such as sending an FP32-sensitive tensor through low-bit.

This separation is the state boundary used by the profile-guided online-control mechanisms in Section~\ref{sec:profile-guided-online-control}. Given that boundary, the next subsection walks through how one training step consumes the interface.

\subsection{Runtime Execution}

At runtime, each training step follows the same sequence:

\begin{enumerate}
\item The model computes gradients normally.
\item DDP forms communication buckets.
\item The PyTorch DDP communication hook uses the workload/model binding to classify bucket contents.
\item Commander admission and the current Supervisor state determine whether the eligible portion may use low-bit or must fall back to FP32.
\item For admitted eligible slices, the hook performs write-side payload materialization: FP32 gradients are encoded as G-Binary or G-Ternary sign/ternary payloads before communication, while rejected or sensitive slices remain on the FP32 path.
\item The endpoint-capacity guard admits the low-bit reducer route only when endpoint service is not expected to become the bottleneck.
\item On the read/response side, the reducer path aggregates the selected payloads with the corresponding FP32 or sign-count semantics and returns the aggregate gradient representation; the training loop then records loss/metric signals for the next Supervisor update.
\end{enumerate}

The important design choice is that the PyTorch DDP communication hook preserves semantic structure below bucket granularity while separating write-side encoding from read/response-side aggregation. This makes the runtime robust to practical DDP bucketization: the system does not need to rely on manual bucket-size tuning to keep sensitive Transformer tensors away from eligible encoder weights, and the reducer-capacity decision remains attached to the aggregation path rather than to compression alone.

The rest of the system description follows this component chain. Section~4 explains how operating profiles, Commander admission, and Supervisor recovery are calibrated. Section~5 explains how runtime bindings and DDP hooks execute those profiles inside mixed buckets. Section~6 explains how reducer-path measurements and endpoint-capacity checks decide whether an admitted low-bit route is architecturally useful.

\section{Operating-Profile Calibration}

Calibration exports an operating profile: a compact runtime input for admitting a low-bit communication route and initializing recovery behavior. Commander uses FP32 warm-up diagnostics and calibrated evidence to admit a candidate mode and model scope; Supervisor uses live training health to choose fallback, cooldown, and readmission after admission. The runtime consumes the profile and its provenance, not the full calibration workflow. This section defines the profile schema, the Commander/Supervisor calibration logic, and the portability boundary; fully zero-shot profile synthesis for unseen workload/model families remains outside the evaluated scope.

\subsection{Operating Profile Schema}

The operating profile is the runtime-facing representation of the calibrated Commander/Supervisor contract. It is a read-only prior: warm-up statistics and live health signals can still change the effective Supervisor state during execution. In the CUSUM detector used by Supervisor~\cite{page1954continuous}, \emph{slack} is the tolerated per-step increase in the smoothed training-health signal before degradation evidence accumulates, and the \emph{threshold} is the accumulated excess that triggers fallback to FP32.

\begin{table}[t]
\caption{Operating-profile calibration summary. Each row records the admissible communication mode and initial recovery settings exported to the runtime; later accuracy-bearing claims are evaluated in Section~7.}
\label{tab:operating-profiles}
\scriptsize
\setlength{\tabcolsep}{3pt}
\begin{tabularx}{\linewidth}{YYYYY}
\toprule
Workload profile & Source & Selected mode & Admission and recovery settings & Calibration/profile evidence \\
\midrule
CIFAR-100/ResNet-18 profile & Vision calibration trace & G-Ternary backbone, FP32 head & admission after epoch 16; slack $3.91\times10^{-3}$; threshold $9.38\times10^{-2}$; one-epoch cooldown and holdout guard & 70.31\% final accuracy, 0.735x FP32 communication, 3/3 true detections, 0 false detections \\
SST-2/DistilBERT profile & Text calibration trace & G-Ternary backbone, FP32 head & post-warm-up admission; slack $2.11\times10^{-4}$; threshold $1.01\times10^{-2}$; one-epoch cooldown; 2x post-recovery guard & 75.27\% final accuracy, 0.372x FP32 communication, 3/3 true detections, 0 false detections \\
\bottomrule
\end{tabularx}
\end{table}

Table~\ref{tab:operating-profiles} records calibration inputs; Sections~5 and~7 report the corresponding execution and validation outcomes. Runtime path counts, route-correctness checks, validation status, and cluster-run metadata remain execution evidence rather than prior state.

\subsection{Commander Admission}

Commander starts in FP32 and uses warm-up diagnostics to decide whether the runtime may admit a candidate low-bit mode, and at what model granularity. Admission is therefore not a single global threshold: full-path low-bit can be rejected while a layer-aware or module-aware route remains admissible for regions such as a convolutional backbone or Transformer encoder, with sensitive regions kept in FP32.

The operating profile records the resulting admission choice as explicit runtime input. For CIFAR-100/ResNet-18, it exports G-Ternary aggregation for the convolutional backbone and FP32 for the classifier head. For SST-2/DistilBERT, it exports a G-Ternary candidate route whose concrete parameter coverage is supplied by the Transformer binding in Section~5. Commander therefore specifies which low-bit route is allowed and which model region it may cover; the online controller later determines how often that allowed route is exercised. Table~\ref{tab:operating-profiles} records the supporting admission evidence.

\subsection{Supervisor Recovery}

The Supervisor uses a calibrated online detector to decide when low-bit execution should fall back to FP32. It monitors a smoothed training-health signal and accumulates degradation evidence with CUSUM. The profile stores initial slack and threshold values, but they are meaningful only relative to the live signal scale; calibration and runtime monitoring therefore use the same smoothed-loss trend.

The calibration summary in Table~\ref{tab:operating-profiles} records the detector scale and recovery settings used to initialize Supervisor. These checks define Supervisor as a recovery controller with fallback, cooldown, and guarded readmission rather than as a post-hoc warning signal. When the detector fires during low-bit execution, the runtime enters FP32 cooldown or readmission guard according to the profile. Online evidence can refine the effective recovery length: near-threshold triggers with small current loss movement can use short recovery, while stronger detector excursions preserve full FP32 fallback. The next subsection describes the evaluated online-control mechanisms that further condition readmission on trigger history and late-run health.

\subsection{Profile-Guided Online Control}
\label{sec:profile-guided-online-control}

Building on the calibration-to-runtime interface in Figure~\ref{fig:calibration-runtime-interface}, profile-guided online control adjusts only run-time effective settings: detector scale, route-use target, cooldown, and readmission. The source operating profile and workload/model binding remain fixed inputs to the run.

The evaluated mechanisms make three bounded moves. First, warm-up loss trends can adjust CUSUM slack and threshold, with \emph{trust-region projection} limiting detector-loosening updates away from the profile prior. Second, warm-up confidence and short \emph{low-bit probe windows} estimate a route-use target; the controller vetoes additional low-bit execution when realized route share exceeds that target plus a guard band. Third, Supervisor recovery chooses effective fallback and readmission behavior from trigger history and health evidence, including repeated-trigger density, near-threshold trigger severity, and late-run health checks. Appendix Table~\ref{tab:app-online-controller-details} records the detailed rules and runtime evidence.

These mechanisms produce runtime state and run evidence; they do not rewrite the source operating profile. Appendix Tables~\ref{tab:app-cifar-controller-variants}, \ref{tab:app-transformer-controller-relaxation}, and~\ref{tab:app-transformer-supervisor-followups} summarize the evaluated controller studies, and Section~7 draws the cross-workload controller lesson.

\subsection{Calibration Cost and Portability}
\label{sec:calibration-portability}

The bounded-control model also defines what can transfer across runs. NEURON-Fabric does not synthesize a new profile or binding from scratch for an unseen workload. Instead, it amortizes a compact offline profile over repeated runs from the same workload/model family, uses a per-model binding to preserve semantic roles, and adapts online only for quantities that the runtime can observe safely during training. Profile fields are revisited when the workload, optimizer, loss scale, or stability target changes; bindings are rebuilt for new model structures; endpoint guards are refreshed when hardware, collective implementation, or topology changes.

The next step is to bind this calibrated control input to an actual distributed-training runtime. Section~5 describes how NEURON-Fabric maps the operating profile onto model structure and DDP communication buckets, including cases where one bucket contains both low-bit-eligible and FP32-sensitive parameters.

\section{Runtime Binding And Bucket Routing}

An operating profile says what kind of low-bit route is admissible, but it does not identify the tensors to which that route applies. Runtime binding supplies this missing structural information. It maps model parameters and DDP bucket contents to roles such as low-bit-eligible backbone tensors, FP32-sensitive head tensors, embeddings, layer norms, biases, and plain FP32 tensors. The PyTorch DDP communication hook then combines the binding with Commander admission, Supervisor state, and endpoint-capacity admission to choose the live route.

Section~2 established the bucketization constraint: DDP buckets are communication units, not model-semantic units. Section~5 turns that constraint into the runtime binding mechanism. A route is dynamic--the same eligible parameter can take FP32 during warm-up, low-bit after Commander admission, FP32 fallback during a Supervisor veto, and low-bit again after cooldown and readmission--but the binding keeps those state changes attached to model semantics rather than to accidental bucket layout. NEURON-Fabric therefore preserves structural roles below bucket granularity: the DDP hook can apply the low-bit estimator to eligible parameter slices while leaving sensitive slices in FP32 inside the same DDP bucket. The rest of this section defines the binding schema and preflight procedure, explains the DDP hook prototype semantics used to validate low-bit route execution, describes concrete workload/model binding instances, and then validates those bindings using route-correctness checks from the DDP hook; Section~7 and Appendix Table~\ref{tab:app-runtime-binding} extend the same route-audit contract to autoregressive language-model runtime checks.

\subsection{Binding Schema And Preflight Procedure}

The binding tells the runtime how each trainable parameter may participate in gradient communication. For each trainable parameter, the binding stores a stable identifier and a semantic role, such as encoder weight, embedding, layer norm, bias, or head. It then records whether that parameter is a low-bit candidate or must remain FP32; for low-bit candidates, it also records the allowed low-bit family and the reason for the assignment. A parameter that is absent from the binding is treated as FP32-only. We use \emph{preflight} to mean the validation pass performed before low-bit execution is enabled. Preflight verifies that the binding covers the expected parameters, that sensitive roles remain FP32, and that the allowed low-bit family matches the operating profile. It is separate from FP32 warm-up: preflight validates the runtime contract, whereas warm-up is part of the training run and supplies current-run diagnostics to Commander and Supervisor. The default is intentionally conservative: an unmapped or structurally sensitive tensor should not enter the low-bit path merely because it shares a DDP bucket with eligible weights. This keeps the runtime claim auditable while still letting model-specific structure expose communication opportunity for backbone or encoder tensors. The same rule applies when expanding low-bit eligibility beyond the currently admitted model regions: newly admitted parameter classes require matching calibration and preflight evidence rather than being enabled by default.

The preflight procedure has two stages. The static stage enumerates model parameters, applies the binding schema, and checks coverage, sensitivity, and expected low-bit share. The executable stage runs the model under DDP with the runtime communication hook enabled, first during FP32 warm-up, then after Commander admission, then under a forced Supervisor veto. This verifies that the same binding drives all runtime states and that DDP bucketization does not erase semantic roles. Appendix Table~\ref{tab:app-runtime-binding} records the current DistilBERT, BERT-base-shaped, real-data Transformer, and Pythia-based autoregressive language-model runtime route evidence.

\subsection{DDP Hook Prototype Semantics}

The binding determines which parameter slices may take the low-bit route; the PyTorch DDP hook prototype defines what that route does during execution. It validates the low-bit route semantics without claiming to be a production packed-sign collective. After backward computation produces FP32 gradients and the controller admits low-bit execution, the worker-side hook performs write-side payload materialization for eligible slices: it records a mean-absolute scale and converts FP32 gradients into dense positive-sign indicators that stand in for packed signs or a ternary payload. Ordinary PyTorch collectives then sum these indicators across workers; the hook applies majority voting, optionally applies the G-Ternary zero-gating pattern, and writes the scaled aggregate direction back into the gradient buffer. This prototype validates routing, fallback, mixed-bucket execution, and optimizer-visible gradient replacement. The modeled gradient-communication traffic values count the same route events using the selected G-Binary or G-Ternary bits-per-element representation, while Section~\ref{sec:reducer} evaluates the reducer path required to realize compact sign aggregation as a production communication mechanism.

\subsection{Workload/Model Binding Instances}

For ResNet/CIFAR, the binding separates the convolutional backbone from the final classifier head. This matches the calibrated profile: full-path low-bit is unsafe, while a G-Ternary backbone with an FP32 head is admissible under guarded Supervisor control.

For DistilBERT/SST-2, the binding is more detailed because Transformer parameters play different architectural roles. The current binding assigns these roles conservatively from model structure and calibration evidence: encoder weight matrices are treated as the main low-bit-eligible region, while embeddings, layer norms, biases, and the sequence-classification head remain FP32. Binding validation maps about 63\% of parameters to low-bit-eligible encoder weights while keeping the sensitive roles in FP32. The static parameter-traffic estimate is about 0.40x FP32, close to the calibrated profile's approximately 0.37x modeled gradient-communication traffic estimate after accounting for runtime route choices under Supervisor control. Appendix Table~\ref{tab:app-runtime-binding} records the exact binding and runtime values together with the validation runs that exercise them. Systematically learning or expanding precision eligibility across other Transformer parameter classes is an important future direction rather than an assumption in the current evaluation.

The binding is loaded at initialization as a separate runtime input rather than as a field inside the operating profile. In the implementation, this structural map is materialized as a versioned binding-configuration file; preflight records its path, identifier, and schema version together with the generated parameter and group maps. The operating profile carries the calibrated control prior, while the binding configuration maps that prior onto concrete model parameters, semantic roles, and DDP bucket slices for a particular model version. A new model version, structural parameter layout, or binding target typically requires a new binding configuration and preflight pass, whereas changes in dataset, topology, or training horizon do not necessarily require rebinding when the model structure is unchanged. The binding remains fixed during a run and is not learned online in the current system. The next step is to verify that this map is honored during execution rather than only recorded as configuration.

\subsection{Runtime-Contract Validation And Route Auditing}

The binding is validated as an executable runtime object, not only as a static configuration. Transformer validation checks exercise the DDP communication hook across FP32 warm-up, Commander admission, and Supervisor fallback. These runs verify that the runtime can load the Transformer binding, admit low-bit on eligible encoder weights after warm-up, and return to FP32 under a Supervisor veto. Appendix Table~\ref{tab:app-runtime-binding} records the corresponding hook-event and route-correctness evidence.

A real-data Transformer validation uses pretrained DistilBERT on SST-2 to verify that the same binding and hook path execute correctly on real text data, not only on the earlier DistilBERT-shaped synthetic-input check. Appendix Table~\ref{tab:app-runtime-binding} also records analogous route-audit checks for the Pythia-based autoregressive language-model path, extending the same runtime contract to a larger model family. Section~7 evaluates the longer Transformer sequence used for the accuracy and modeled gradient-communication traffic claims.

The hook audits these validation runs by recording route counts and route-correctness violations. Route counts summarize how often tensors took the low-bit route, FP32 fallback route, or plain FP32 route. Using the path-count notation introduced in Section~\ref{sec:calibration-runtime-interface}, \texttt{\detokenize{L884/F3915/P0}} denotes 884 low-bit events, 3915 fallback events, and 0 plain-FP32 events. These counts are not policy inputs. They are execution evidence that the runtime actually followed the intended route.

A route-correctness violation occurs when the observed runtime route disagrees with the route implied by the binding and Supervisor state. For example, a tensor marked FP32-sensitive should not be sent through the low-bit estimator, and a low-bit-eligible tensor under an active Supervisor veto should take FP32 fallback rather than low-bit. The check compares the hook's recorded route against this expected role-and-state decision. The current CIFAR/ResNet training runs with live Commander and Supervisor control, Transformer runtime validations, and autoregressive language-model runtime checks report zero route-correctness violations; the runtime-binding rows in Appendix Table~\ref{tab:app-runtime-binding} record the corresponding route-audit evidence.

This section supports a systems claim about execution, not just configuration. NEURON-Fabric binds the operating profile to model structure, preserves that binding inside mixed DDP buckets, and records the actual runtime route taken by the hook. These checks establish that low-bit routes can be selected correctly at runtime. The next question is architectural: once a route is numerically safe and structurally valid, when does the communication system actually benefit from sending it through a low-bit reducer rather than through FP32 communication? Section~6 answers this question by defining the reducer co-design problem and the reducer-admission region used by the runtime guard.

\section{Reducer Co-Design And Admission Region}
\label{sec:reducer}

Section~2 identified reducer capacity as a separate systems constraint: reducing gradient payloads helps only when the communication substrate can aggregate the compact representation without creating a new bottleneck. This section turns that constraint into a guarded reducer-admission problem. The operating profile and runtime binding decide when a low-bit route is safe for the workload and model; reducer admission decides whether that safe route is also useful for the current topology, endpoint service rate, and worker fan-in.

The evidence below answers two questions. First, the reducer path must realize the runtime contract established in Section~5: the runtime must select the intended path, fall back when the controller or endpoint guard rejects low-bit, and return correct aggregate results. Second, the route must be profitable: the reducer path should be admitted only when it is expected to reduce gradient-communication cost relative to the matched FP32 path.

\subsection{Reducer Model and Reference Paths}

The reducer studied in this paper is a hierarchical sign-count reducer, following the broader systems direction of moving reduction closer to network or endpoint substrates~\cite{graham2016sharp,sapio2021switchml,gebara2021panama}: it aggregates compact worker updates through node-local counting and inter-node partial-count exchange before returning a synchronized aggregate direction. Thus, the model and optimizer (e.g., SGD or AdamW) still see a normal synchronized gradient buffer, while the communication substrate decides whether compact writes can be counted and returned efficiently enough to justify the low-bit route.

The reducer analysis compares two reference communication paths. The first is the FP32 path, which uses measured PyTorch/NCCL all-reduce on numeric gradient tensors~\cite{li2020pytorchddp}. The second is the stock-collective baseline for sign-count aggregation when no custom reducer exists, implemented here as packed-sign all-gather followed by local unpack-and-count work at each worker.

We do not assume a specific NCCL ring, tree, or direct algorithm; instead, communication times for the FP32 reference and stock packed-sign baseline are taken from measured PyTorch/NCCL payload-time curves for all-reduce and all-gather, respectively. Standard all-reduce sums numeric tensor entries, whereas sign-count aggregation over bit-packed updates requires recovering per-coordinate sign counts rather than numerically adding packed machine words. The proposed reducer is modeled semantically as node-local sign counting followed by inter-node partial-count exchange and result return.

\subsection{Bucket-Communication Replay}

Replay combines measured PyTorch/NCCL payload curves with real DDP bucket schedules. Across six two-node, four-node, and longer-trace cases, the low-bit reducer model reduces replayed bucket communication by 5.19--47.22x relative to matched FP32; Appendix Table~\ref{tab:app-reducer-replay} records the cases. This establishes communication opportunity, but not by itself that stock collectives realize the sign-count reducer semantics required by the low-bit route.

\subsection{Why Packed Signs Over Stock Collectives Are Not Enough}

\begin{figure}[t]
\centering
\includegraphics[width=0.82\linewidth]{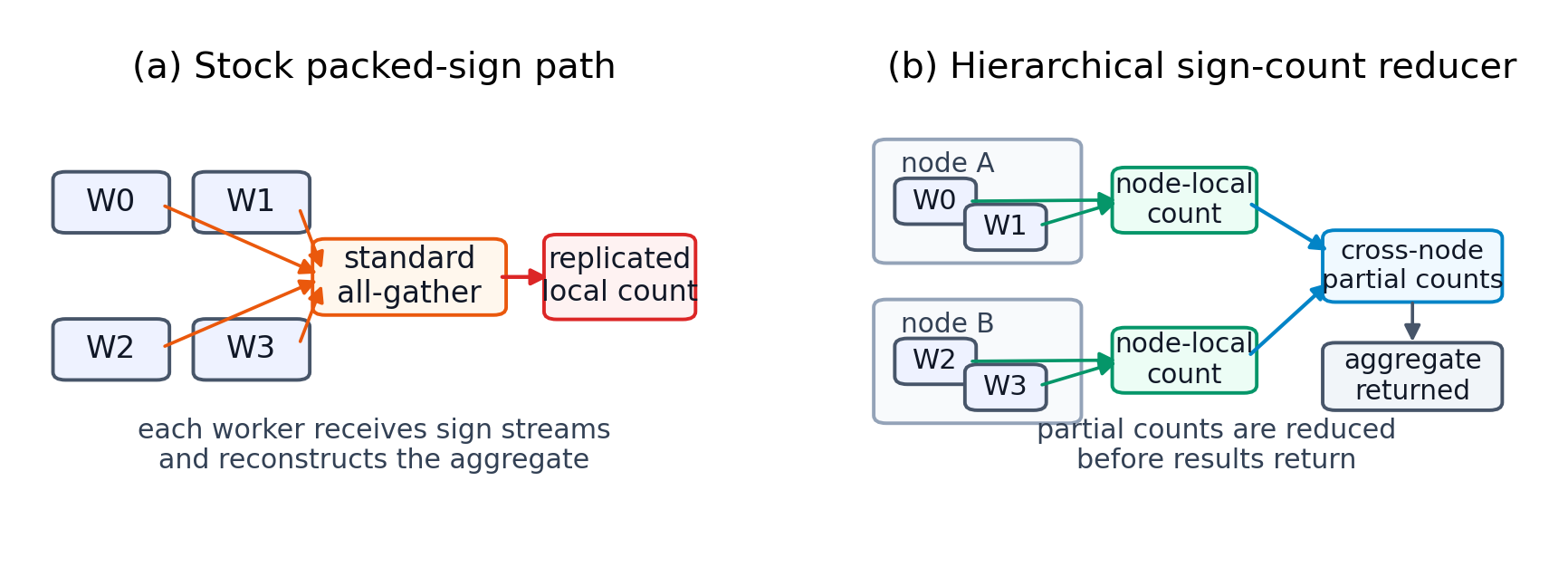}
\caption{Communication-path contrast for sign-count aggregation. Stock packed-sign transport exchanges sign streams and repeats counting at each worker, while the hierarchical reducer performs node-local counting, cross-node partial-count exchange, and aggregate return.}
\Description{Two-panel schematic contrasting stock packed-sign transport, where workers exchange sign streams and reconstruct the aggregate locally, with a hierarchical sign-count reducer, where node-local counting is followed by cross-node partial-count exchange and aggregate return.}
\label{fig:reducer-topology-contrast}
\end{figure}

Whether that communication opportunity is useful depends on where aggregation occurs. Stock packed-sign transport reduces the payload representation, but aggregation remains replicated: each worker receives sign streams and reconstructs the same count locally. The hierarchical reducer instead performs node-local sign counting, exchanges partial counts, and returns the aggregate direction. The communication pattern is therefore partial-count reduction plus result return, rather than replicated packed-sign transport followed by local counting. Figure~\ref{fig:reducer-topology-contrast} makes this distinction explicit.

A grouped-timeline replay confirms the semantic distinction. Hierarchical sign-count aggregation is far below stock packed-sign transport because it avoids replicated local counting, but it can still lose to grouped FP32 when endpoint work dominates. Appendix Table~\ref{tab:app-reducer-support} records the detailed timing comparison. This motivates an endpoint-capacity guard: low-bit reduction should be admitted when it is beneficial relative to FP32, not merely because it is better than stock packed-sign transport.
\subsection{Endpoint-Capacity Guard}

The \emph{reducer-admission region} is the set of bucket, topology, network, and endpoint-service conditions under which hierarchical sign-count reduction is expected to reduce bucket communication time relative to FP32. The endpoint-capacity guard admits a candidate low-bit bucket only when the estimated hierarchical-reducer time is lower than the matched FP32 time for the same bucket size, topology, local fan-in, endpoint service rate, and network bandwidth. Otherwise, the bucket remains on the FP32 path.

\begin{figure}[t]
\centering
\includegraphics[width=\linewidth]{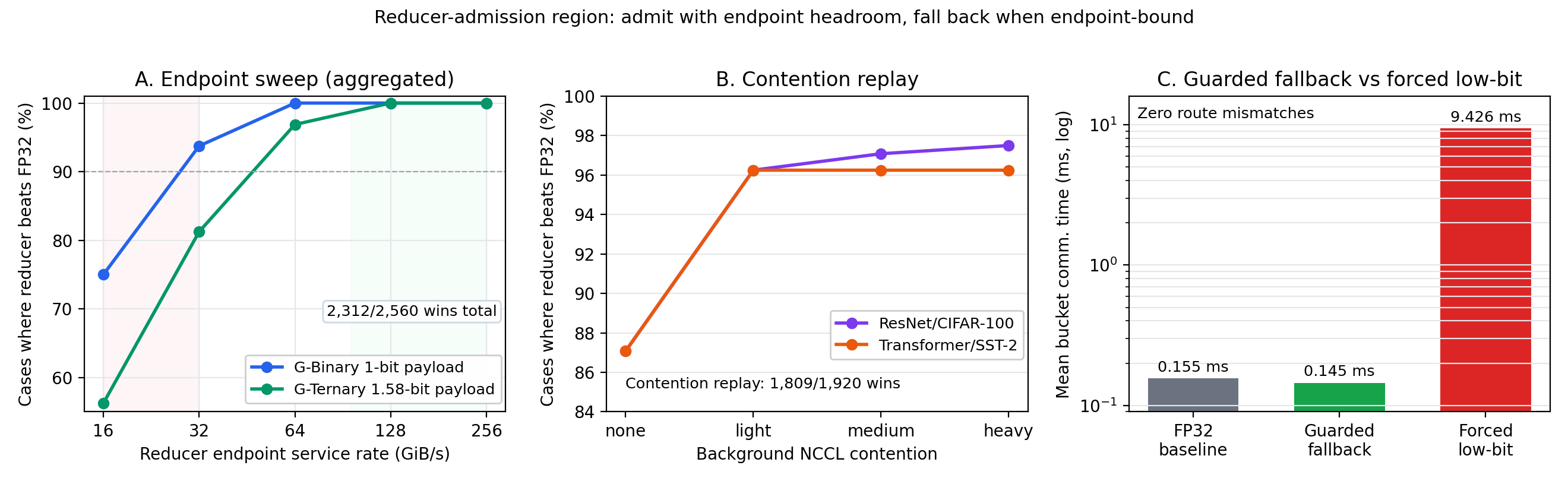}
\caption{Reducer-admission evidence. (A) Aggregated endpoint-capacity sweep. (B) Contention replay under background NCCL traffic. (C) Guarded fallback versus forced low-bit in an endpoint-bound case. Together, the panels delimit the reducer-admission region.}
\Description{Three-panel reducer evidence figure with an aggregated endpoint-capacity sweep, contention replay under background NCCL traffic, and guarded fallback versus forced low-bit in an endpoint-bound case.}
\label{fig:reducer-claim-boundary}
\end{figure}

The sweep results give two lessons. First, fast networks do not make low-bit routing irrelevant; they make endpoint service capacity more important. Across the topology and capacity sweeps, the hierarchical reducer wins most modeled configurations, while losses concentrate under fast networks, high fan-in, and low endpoint service rate. Panel~A of Figure~\ref{fig:reducer-claim-boundary} compresses the capacity sweep by plotting the winning fraction for G-Binary and G-Ternary as endpoint service rate changes; Appendix Table~\ref{tab:app-reducer-support} records the sweep dimensions and counts.

Second, contention increases the value of payload reduction without removing the endpoint boundary. Panel~B applies measured background NCCL slowdown factors to the captured bucket schedules and recomputes whether the low-bit reducer remains faster than matched FP32. Panel~C fixes an endpoint-bound corner: with the guard enabled, the runtime falls back near FP32; when the guard is disabled, forced low-bit queues at the endpoint and loses. Thus the guard is part of the runtime route semantics, not only an offline modeling conclusion.

Executable reduction checks complement the modeled admission region. On a two-node, eight-GPU Narval configuration, distributed bucket executions in reference sign-count, G-Binary, and G-Ternary modes return zero aggregate-output mismatches, and the local counting stage also returns zero mismatches for the tested logical-bucket groups. Detailed reducer-admission evidence, including the local-count component check, is reported in Appendix Table~\ref{tab:app-reducer-support}.

Section~7 then turns to the training question: whether operating profiles, Supervisor recovery, and per-parameter DDP routing can exercise admitted low-bit routes while preserving model accuracy across vision and Transformer workloads.

\section{Closed-Loop Cluster Evaluation}

This section evaluates closed-loop runtime behavior with the prototype PyTorch hook, including accuracy-bearing training runs, matched FP32 anchors, and scale-out runtime checks. It separates this runtime/control validation from production reducer-throughput evaluation: the runs below test whether a calibrated operating profile can configure the DDP communication hook, whether Supervisor fallback recovers accuracy that static low-bit loses, and whether the same binding/hook interface carries to Transformer workloads, BERT-style encoders, and autoregressive language-model runtime checks.

The evaluation follows a dependency chain. CIFAR/ResNet establishes the accuracy--route-share tradeoff and defines the target accuracy region, where successful runs keep final top-1 at or above 70\% with no route-correctness violations. Transformer runs then test transfer through the same profile, binding, and hook interface. Matched FP32 baselines and target thresholds serve as evaluation anchors, not as runtime inputs to the controller.

\subsection{Evaluation Setup and Scope}

All accuracy-bearing training and route-correctness runs execute through PyTorch DDP and NCCL on A100 GPU nodes in the Narval cluster. The real training runs cover CIFAR-100/ResNet-18~\cite{krizhevsky2009cifar,he2016resnet} and SST-2~\cite{socher2013sentiment,wang2019glue} Transformer workloads, including pretrained DistilBERT~\cite{sanh2019distilbert} and BERT-base~\cite{devlin2019bert} follow-ups, over one-, two-, and four-node deployments. Separate short-horizon autoregressive language-model runtime validations use Pythia-1.4B and Pythia-6.9B~\cite{biderman2023pythia} on WikiText-103~\cite{merity2017pointer} with full-parameter DDP over four nodes. Matched FP32 runs anchor the loss and perplexity comparisons. Reported modeled gradient-communication traffic is normalized to matched FP32 execution from the realized runtime route counts; it is not an end-to-end training-throughput measurement. Appendix Table~\ref{tab:app-eval-environment} summarizes the measurement scope so that measured accuracy, modeled gradient-communication traffic, replayed reducer admission, and reducer-component microbenchmarks are not conflated.

The remaining subsections follow this chain: vision tradeoff, Transformer and autoregressive language-model runtime evidence, and cross-workload controller interpretation.

\subsection{CIFAR/ResNet Accuracy--Route-Share Tradeoff}

The CIFAR/ResNet tradeoff uses the Section~4 operating profile and the Section~5 binding and route-audit path. Table~\ref{tab:vision-frontier} reports an ordered set of operating points that separates accuracy preservation from realized low-bit route share. The FP32 reference establishes the accuracy anchor, while static layer-aware low-bit execution isolates the failure mode: the route can be selected consistently, yet convergence collapses when all eligible traffic is forced onto the low-bit path. Guarded policies then quantify how much low-bit route share can be admitted under Supervisor control while remaining in the target accuracy region. We distinguish final from best top-1 because a run that reaches the target only transiently is not treated as stable. A point is stable if it meets the target with nonzero low-bit routing, boundary if it only transiently reaches the target or sits close to the threshold, and unsafe if final accuracy falls below the target. The Paths column uses the Section~\ref{sec:calibration-runtime-interface} path-count notation \texttt{L/F/P}.

\begin{table}[t]
\caption{Vision accuracy--route-share tradeoff. Static low-bit exposes the failure mode; guarded Supervisor control recovers most FP32 accuracy while keeping low-bit route share inside the target accuracy region.}
\label{tab:vision-frontier}
\scriptsize
\setlength{\tabcolsep}{3pt}
\begin{tabularx}{\linewidth}{YYYYYY}
\toprule
Evidence point & Topology & Final / best top-1 & Low-bit & Paths & Interpretation \\
\midrule
FP32 reference & 2 nodes x 4 GPUs & 73.39 / 73.51\% & 0.00\% & \pathcounts{0}{0}{4799} & Full-precision target. \\
Static layer-aware low-bit & 2 nodes x 4 GPUs & 16.88 / 16.90\% & 100.00\% & \pathcounts{4799}{0}{0} & Correct route, failed convergence. \\
Calibrated Supervisor & 2 nodes x 4 GPUs & 70.59 / 70.59\% & 18.42\% & \pathcounts{884}{3915}{0} & Stable target-region point with zero route-correctness violations. \\
Best tradeoff point & 2 nodes x 4 GPUs & 70.14 / 70.21\% & 21.54\% & \pathcounts{1034}{3765}{0} & Highest low-bit clean point that remains in the target region. \\
Boundary point & 2 nodes x 4 GPUs & 68.83 / 70.18\% & 24.00\% & \pathcounts{1152}{3647}{0} & Best checkpoint reaches the target region, but final accuracy falls below it. \\
Unsafe region & 2 nodes x 4 GPUs & 67.91 / 67.91\% & 32.00\% & \pathcounts{1536}{3263}{0} & Final accuracy remains below the target region. \\
Four-node transfer & 4 nodes x 4 GPUs & 70.23 / 71.26\% & 24.00\% & \pathcounts{1152}{3647}{0} & Closed-loop behavior survives topology movement. \\
\bottomrule
\end{tabularx}
\end{table}

\begin{figure}[t]
\centering
\includegraphics[width=\linewidth]{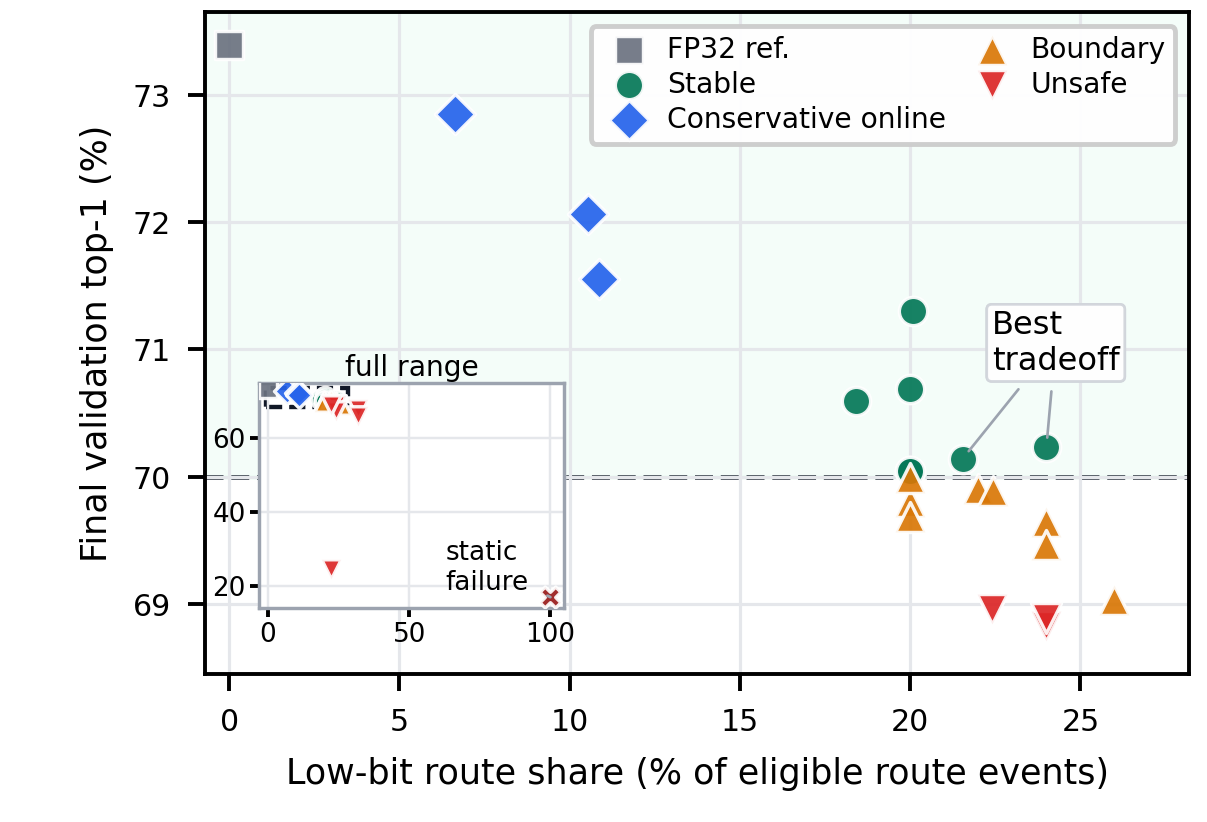}
\caption{CIFAR/ResNet accuracy--route-share tradeoff. Stable and conservative online
points remain inside the target accuracy region, boundary points sit near the
target threshold, and unsafe points fall below it. The highlighted best-tradeoff
point marks the evaluated point with the strongest combination of final
accuracy and low-bit route share.}
\Description{Scatter plot relating low-bit route share to final accuracy, with guarded policies near the safe region and aggressive policies beyond it.}
\label{fig:closed-loop-frontier}
\end{figure}

Figure~\ref{fig:closed-loop-frontier} visualizes the same evidence as an accuracy--route-share frontier. The target threshold marks the minimum acceptable final accuracy for this workload. Policies around 20--22\% low-bit route share remain in the stable region; the 24\% point reaches the target only transiently and therefore marks the current boundary; and the 32\% point illustrates over-admission. The four-node transfer repeats the selected route-share regime after topology movement and remains above the target with zero route-correctness violations, supporting the claim that the evaluated control path is not tied to the original two-node deployment.

\subsection{Transformer Runtime And Transfer Evidence}

Section~5 established that the Transformer binding and DDP hook preserve model semantics inside mixed buckets. The accuracy-bearing Transformer checks keep that executable path and report validation accuracy, low-bit route share, veto share, and modeled gradient-communication traffic. The main 12-epoch DistilBERT validation reaches 87.9\% mean accuracy, 29.1\% mean low-bit route share, and 0.824x modeled gradient-communication traffic relative to FP32. Controlled topology, dataset/step-scale, combined 2-node, batch-side, and BERT-base follow-ups remain within the evaluated Transformer boundary, with the pretrained BERT-base run reaching 90.1\% mean accuracy and 0.777x modeled gradient-communication traffic. Appendix Table~\ref{tab:app-transformer-support} reports the exact values and matched FP32 references.

The next validation moves beyond encoder-based text classification. It uses autoregressive language models in a continued-pretraining setting to test whether the same route-audit path also executes on decoder-only Transformer workloads. Table~\ref{tab:llm-runtime-scale} reports two full-parameter DDP continued-pretraining runs on WikiText-103 raw text. These checks evaluate whether the existing control and routing interface remains executable at autoregressive language-model scale, rather than defining a new Pythia-specific calibration profile. The matched FP32 rows make the loss and perplexity comparison explicit.

\begin{table}[t]
\caption{Autoregressive language-model scale-out runtime validation with matched FP32 anchors. Both runs use full-parameter DDP with the prototype PyTorch hook; MGTR is computed from realized route counts.}
\label{tab:llm-runtime-scale}
\tiny
\setlength{\tabcolsep}{2pt}
\begin{tabularx}{\linewidth}{YYYYYYYYY}
\toprule
Run & Model scale & Deployment & Horizon & Validation loss (low-bit / FP32) & Perplexity (low-bit / FP32) & Low-bit share & MGTR & Route violations \\
\midrule
Pythia-1.4B & 1.41B parameters & 4 nodes x 4 A100 GPUs & 96 steps, sequence length 512 & 2.457 / 2.458 FP32 & 11.67 / 11.69 FP32 & 30.21\% & 0.758x & 0 \\
Pythia-6.9B & 6.86B parameters & 4 nodes x 4 A100 GPUs & 24 steps, sequence length 128 & 2.937 / 2.939 FP32 & 18.86 / 18.90 FP32 & 29.17\% & 0.743x & 0 \\
\bottomrule
\end{tabularx}
\end{table}

These results extend the route-audit path from encoder-based text classification to decoder-only language-model training while keeping quality comparisons anchored by matched FP32 references. They establish that the same profile, binding, hook, and route-correctness interface remains executable across the evaluated model families and scales. The remaining question is what these runs reveal about controller design: when online adaptation should admit more low-bit traffic, when it should remain conservative, and when Supervisor recovery should return execution to FP32.

\subsection{Controller Evidence Across Workloads}

Across workloads, the controller evidence changes the objective from maximizing low-bit residence to regulating low-bit exposure. The CIFAR frontier shows that route correctness alone is insufficient: aggressive low-bit admission can execute the intended path while leaving the target accuracy region. The Transformer sequence shows the complementary case: a transferred profile can be overly conservative unless recovery and readmission respond to observed run behavior.

The cross-workload lesson is that useful low-bit communication depends on controlled exposure, not on a static compression decision. Warm-up and probe evidence can reveal additional opportunity, but Supervisor recovery must distinguish benign fluctuations from degradation and must avoid shrink-biased target updates that suppress useful low-bit traffic. Detailed controller evidence is reported in the appendix; the main text uses it to support a single design conclusion: profile-guided online control should adapt readmission and recovery behavior within the calibrated workload/model boundary. Appendix Tables~\ref{tab:app-online-controller-details}, \ref{tab:app-cifar-controller-variants}, \ref{tab:app-transformer-controller-relaxation}, \ref{tab:app-transformer-supervisor-followups}, and~\ref{tab:app-transformer-support} provide the supporting evidence.

\section{Related Work}

\subsection{Gradient Compression and Low-Bit Training}

Gradient compression has a long systems and optimization lineage. Early 1-bit stochastic gradient descent (SGD), QSGD, TernGrad, Deep Gradient Compression, Error-Feedback SGD, and PowerSGD explore stochastic quantization, ternary gradients, sparsification, residual/error feedback, and low-rank compression \cite{seide2014onebit,alistarh2017qsgd,wen2017terngrad,lin2018dgc,karimireddy2019error,vogels2019powersgd}. These methods reduce the bytes exchanged by a communication primitive while preserving enough optimization signal for convergence. NEURON-Fabric shares that motivation but places the low-bit path behind a runtime controller: low-bit aggregation is admitted, monitored, withdrawn, and readmitted rather than installed as a static replacement for FP32 all-reduce.

Binary and ternary low-bit training methods show that selected neural-network computations can tolerate heavily quantized arithmetic \cite{courbariaux2015binaryconnect,rastegari2016xnor,wen2017terngrad}. They motivate the sign-count and ternary-gating arithmetic used by G-Binary and G-Ternary, but NEURON-Fabric applies these ideas only to gradient communication payloads. The model architecture, local forward/backward computation, and optimizer update remain unchanged. The operating profile selects the candidate low-bit route, while the workload/model binding keeps classifier heads, embeddings, layer norms, biases, and other sensitive regions on FP32 even when eligible backbone or encoder weights use low-bit communication. This distinction is especially important for Transformer workloads, where mixed DDP buckets can otherwise force whole-bucket fallback.

\subsection{Adaptive Training and Cluster Control}

Training systems already use runtime evidence to protect useful progress. Early stopping monitors validation behavior \cite{prechelt1998early}; gradient clipping responds to unstable gradient norms \cite{pascanu2013difficulty}; mixed-precision training adjusts loss scaling when numerical overflow appears \cite{micikevicius2018mixed}; and bandit-style hyperparameter search allocates resources toward promising configurations \cite{li2017hyperband}. NEURON-Fabric follows the same control pattern but uses a different actuator: the runtime switches eligible gradient communication between FP32 and low-bit routes as health signals and route usage evolve. Its CUSUM-style detector draws on sequential change detection \cite{page1954continuous}, but the response is a precision-route transition with fallback and readmission semantics.

GPU-cluster systems use related control ideas at the job and resource-management layers. Gandiva, Tiresias, Pollux, Optimus, Sia, and Shockwave use runtime observations for time-slicing, migration, scheduling, placement, resource allocation, goodput optimization, or fairness-aware planning \cite{xiao2018gandiva,gu2019tiresias,qiao2021pollux,peng2018optimus,subramanya2023sia,zheng2023shockwave}. NEURON-Fabric moves the control boundary inside the gradient communication path. Its actuator is not cluster placement, job scheduling, or batch-size selection; it is the low-bit versus FP32 route for eligible gradients, subject to model semantics, training health, and reducer capacity.

\subsection{Distributed Training Runtimes and Bucketization}

Distributed training runtimes expose communication hooks, bucketization policies, collective libraries, and topology-aware scheduling. Parameter-server systems, TensorFlow, PyTorch DDP, Poseidon, Blink, and ByteScheduler demonstrate how runtime substrates and tensor-scheduling policies shape distributed-training performance \cite{li2014parameter,abadi2016tensorflow,li2020pytorchddp,zhang2017poseidon,wang2020blink,peng2019bytescheduler}. NEURON-Fabric builds on this substrate but treats bucketization as a semantic hazard. A DDP bucket is a communication unit, not a model-structure unit, so the runtime needs a binding layer that maps parameters to roles before the hook chooses low-bit or FP32 routes.

\subsection{Reducers, In-Network Aggregation, and Endpoint Guards}

In-network aggregation and hierarchical collective designs move reduction work closer to the network or reducer substrate. SHArP offloads reduction into InfiniBand switch hardware, SwitchML co-designs programmable-switch aggregation with end-host protocols, PANAMA studies shared-cluster in-network aggregation with fairness concerns, and CXL-style coherent accelerator and memory fabrics provide another setting in which endpoint service and fabric bandwidth must be modeled together \cite{graham2016sharp,sapio2021switchml,gebara2021panama,cxl2022spec}. These systems motivate reducer-side execution. NEURON-Fabric focuses on the admission decision around such a reducer: low-bit communication is selected only when the current bucket, topology, contention regime, endpoint-capacity model, and training-health state indicate that the route should be both useful and safe.

\section{Limitations And Future Work}

NEURON-Fabric makes a bounded systems claim. It shows that safe low-bit gradient aggregation requires a calibrated operating profile, workload/model binding, online recovery control, bucket-aware routing, and guarded reducer admission. It does not claim universal zero-shot low-bit training, automatic binding for every model family, arbitrary batch or optimizer invariance, or production end-to-end speedup from the current PyTorch hook.

The evidence covers CIFAR-100/ResNet-18 and SST-2 Transformer paths, including DistilBERT, BERT-base-shaped binding checks, seed and training-horizon checks, topology movement, dataset/step-scale movement, and narrow batch side points. It also includes short-horizon autoregressive language-model runtime checks with Pythia-1.4B and Pythia-6.9B, anchored by matched FP32 loss and perplexity references. These checks validate scale-out route execution, not arbitrary language-model generality. A new workload/model family still needs an operating profile, binding adapter, and preflight check; NEURON-Fabric currently amortizes calibration within evaluated families rather than synthesizing a universal zero-shot profile for arbitrary workloads.

The controller is also only partially online. Commander and Supervisor use warm-up, probe, route-accounting, and live health evidence to change effective route-use targets, cooldown, readmission, and recovery length, but the calibrated operating profile remains the prior. The current controller already includes preliminary phase-related signals through normalized progress, live health evidence, trigger history, and noise-aware recovery. However, these mechanisms do not yet form an explicit phase-aware readmission policy. Future work should make Commander admission and Supervisor recovery more evidence-adaptive, including adaptive guard-band selection, hysteresis-based target updates, and trigger classification that separates noise-like events from true degradation. A natural next step is phase-aware recovery: stable late-stage plateau oscillations may permit more aggressive low-bit readmission than early training, while sustained late-run degradation should still trigger FP32 fallback.

Future work should also automate binding discovery and rebinding for unseen model structures rather than requiring a manually prepared binding-configuration file and preflight pass for each new structural target, move more workload/model-specific profile construction into warm-up and short-probe calibration, broaden optimizer, batch-size, topology, and workload coverage, and implement the low-bit route as a production custom collective or CXL-side reducer measured inside full training runs.

\section{Conclusion}

NEURON-Fabric turns low-bit gradient aggregation from a static compression choice into a controlled runtime path. Across the evaluated vision, Transformer, and autoregressive language-model workloads, the system admits low-bit communication through a calibrated operating profile, binds it to model structure, preserves semantics below DDP bucket granularity, monitors health through online recovery control, and guards admission against reducer endpoint capacity. The Pythia-1.4B and Pythia-6.9B scale-out checks show that the same hook/control path executes at billion-parameter autoregressive language-model scale with matched short-horizon FP32 loss/perplexity anchors. The results validate profile-guided execution, mixed-bucket routing, fallback/readmission behavior, accuracy-bearing low-bit route share, scale-out runtime execution, and the reducer-admission conditions under which hierarchical sign-count aggregation can lower gradient-communication cost.

\clearpage
\appendix
\section{Supporting Evidence Tables}
\label{app:evidence}

This appendix records supporting evidence that would otherwise interrupt the main text. It is organized by evidence type rather than by claim summary: controller-mechanism details, CIFAR/ResNet controller variants, runtime binding and route checks, reducer-admission support, and Transformer/autoregressive language-model validation. The main body states the system claim and representative results; the tables below keep the underlying configurations and outcomes auditable without restating the narrative.

\subsection{Calibration And Controller Support}

Table~\ref{tab:app-online-controller-details} expands the implementation details behind the compact controller equations in Section~4.4. These rules are runtime mechanisms: they consume the operating profile and online observations, then report effective settings and route evidence in the run summary rather than rewriting the source profile.

\begin{table}[t]
\caption{Runtime inputs, implemented rules, and recorded evidence for the evaluated online-control mechanisms.}
\label{tab:app-online-controller-details}
\scriptsize
\setlength{\tabcolsep}{3pt}
\begin{tabularx}{\linewidth}{YYYY}
\toprule
Mechanism & Runtime inputs & Implemented rule & Recorded evidence \\
\midrule
Trust-region detector calibration & Profile CUSUM settings; FP32 warm-up loss deltas; confidence & Estimate run-local slack/threshold and bound detector-loosening updates within a trust region & Effective detector settings, confidence, projection flag, path counts \\
Route-use budget guard & Target low-bit share; realized share; margin; guard interval & Veto additional low-bit execution when realized route use exceeds the target plus guard margin & Target route share, realized route share, guard-trigger events \\
Adaptive budget windows & Window triggers, CUSUM maximum, loss delta, low-bit share & Increase the route-use target after healthy windows and decrease it after unhealthy or over-budget windows & Initial/final target, budget-update events, window health \\
Late-phase health guard & Training progress, learning-rate fraction, late-window health ratios, healthy-window references & Gate late-stage route growth using progress, learning-rate state, and healthy-window references & Late-phase guard events, learned references, final low-bit share \\
Probe-guided Transformer control & Warm-up deltas, probe loss increase, trigger density, remaining horizon & Derive route-use target, cooldown, and readmission window; apply a three-point guard band & Learned target, trigger density, guard band, cooldown, validation accuracy \\
Noise-aware Transformer recovery & Trigger magnitude, current loss movement, effective detector settings & Use short recovery for near-threshold triggers and retain full fallback for larger excursions & Short-recovery count, cooldown behavior, accuracy, low-bit share, MGTR \\
\bottomrule
\end{tabularx}
\end{table}

Table~\ref{tab:app-cifar-controller-variants} reports the CIFAR/ResNet controller sequence used to motivate bounded online adaptation. Each row preserves the same workload family, binding, and route-checking path while changing the online-control rule, so the table should be read as a controlled controller study rather than as independent workload evidence.

\begin{table}[t]
\caption{CIFAR/ResNet controller sequence for profile-guided online control.}
\label{tab:app-cifar-controller-variants}
\scriptsize
\setlength{\tabcolsep}{3pt}
\begin{tabularx}{\linewidth}{YYYY}
\toprule
Control rule & Outcome & Interpretation & Controller implication \\
\midrule
Warm-up-only calibration & 32.00\% low-bit route share; 65.85\% final top-1 & Warm-up diagnostics alone over-admit low-bit traffic and leave the target-accuracy region & Online admission needs recovery and exposure bounds after warm-up. \\
Trust-region detector projection & 22.42\% low-bit route share; 68.96\% final top-1 & Bounding detector movement reduces the warm-up-only failure but does not restore target-region accuracy & Detector projection is useful but insufficient without route-use and health constraints. \\
Auto-discovered route-use budget & 6.63\% low-bit route share; 72.85\% final top-1 & A conservative route-use budget preserves accuracy but leaves communication opportunity unused & Automatic budget discovery is safety-favorable but can be overly conservative. \\
Smoothed budget adaptation & 20.00\% low-bit route share; 69.68\% final top-1 & Smoothing recovers route share near the fixed-profile operating point but remains vulnerable to late-run accuracy loss & Route-use adaptation also needs late-stage health gating. \\
Normalized late-phase health guard & 10.54\% low-bit route share; 72.06\% final top-1; zero route-correctness violations & Late-phase health checks restore final stability on the same checked routing path & Late-stage growth should be conditioned on progress and health evidence. \\
Healthy-window reference relaxation & 10.88\% low-bit route share; 71.55\% final top-1; zero route-correctness violations & A current-run healthy-window reference preserves stability while reducing CIFAR-specific constants & Runtime-derived health references can replace part of the hand-tuned controller state. \\
\bottomrule
\end{tabularx}
\end{table}

\subsection{Runtime Binding And Routing Support}

Table~\ref{tab:app-runtime-binding} records the runtime-binding checks introduced in Section~5. The rows separate static binding coverage from executable route audits: the former checks whether model parameters are assigned to the intended semantic roles, while the latter checks whether the DDP hook honors those roles during warm-up, admission, Supervisor veto, and fallback states.

\begin{table}[t]
\caption{Runtime binding and route-audit evidence.}
\label{tab:app-runtime-binding}
\scriptsize
\setlength{\tabcolsep}{3pt}
\begin{tabularx}{\linewidth}{YYYY}
\toprule
Runtime check & Scope & Evidence recorded & What it validates \\
\midrule
Vision binding check & ResNet/CIFAR backbone-head split & Backbone tensors are low-bit eligible; the classifier head remains FP32; live closed-loop runs report zero route-correctness violations & The binding preserves model roles for the vision workload. \\
Transformer static binding check & DistilBERT/SST-2 parameter map & About 63\% of parameters map to low-bit-eligible encoder weights; embeddings, layer norms, biases, and classifier head remain FP32 & The Transformer binding exposes encoder opportunity while keeping sensitive roles on FP32. \\
Mixed-bucket execution check & DistilBERT-shaped DDP validation & FP32 warm-up, Commander admission, Supervisor veto, and fallback execute through the same hook; 704 hook events are recorded & The hook can preserve semantic roles below DDP bucket granularity. \\
Real-data Transformer route audit & Pretrained DistilBERT on SST-2 text & 86.93\% validation accuracy; 12,704 hook events; 288 mixed-parameter low-bit bucket events; zero route-correctness violations & The same binding and hook path runs on real text data before the longer controller sequence. \\
BERT-style binding follow-up & BERT-base-shaped validation and pretrained BERT-base run & BERT-base-shaped validation passes; pretrained BERT-base reaches 90.06\% mean validation accuracy with 30.27\% mean low-bit route share & The binding path is not hard-coded to DistilBERT parameter names. \\
Autoregressive language-model route audit & Pythia-1.4B and Pythia-6.9B full-parameter DDP validations & Contract status passes; zero route-correctness violations; Pythia-1.4B records 33,408 low-bit bucket events and Pythia-6.9B records 10,752 & The same route-audit contract executes at billion-parameter autoregressive LM scale. \\
\bottomrule
\end{tabularx}
\end{table}

\subsection{Reducer Admission Support}

Table~\ref{tab:app-reducer-replay} records the six payload-replay cases behind the main-text bucket-communication ranges in Section~6. Table~\ref{tab:app-reducer-support} then summarizes the reducer-admission evidence chain, including the local-count component check. This split keeps the detailed payload replay separate from the broader reducer-admission overview.

\begin{table}[t]
\caption{Payload-replay cases underlying the main-text bucket-communication ranges in Section~6.}
\label{tab:app-reducer-replay}
\tiny
\setlength{\tabcolsep}{3pt}
\begin{tabularx}{\linewidth}{llXrrr}
\toprule
Experiment & Topology & Workload & FP32 replay & Low-bit replay & Tail reduction \\
\midrule
Exp29f & 2 nodes & ResNet/CIFAR-100 & 2.85 ms & 0.55 ms & 5.19x \\
Exp29f & 2 nodes & SST-2/Transformer & 11.53 ms & 0.68 ms & 17.05x \\
Exp30c & 4 nodes & ResNet/CIFAR-100 & 4.34 ms & 0.49 ms & 8.87x \\
Exp30c & 4 nodes & SST-2/Transformer & 17.44 ms & 0.70 ms & 24.85x \\
Exp31a & 4 nodes, longer trace & ResNet/CIFAR-100 & 12.57 ms & 0.75 ms & 16.85x \\
Exp31a & 4 nodes, longer trace & SST-2/Transformer & 38.94 ms & 0.83 ms & 47.22x \\
\bottomrule
\end{tabularx}
\end{table}

\begin{table}[t]
\caption{Reducer-admission evidence chain. Detailed payload-replay cases are reported in Table~\ref{tab:app-reducer-replay}; local-count component timings are summarized here because they are small in scope.}
\label{tab:app-reducer-support}
\scriptsize
\setlength{\tabcolsep}{3pt}
\begin{tabularx}{\linewidth}{YYYY}
\toprule
Reducer question & Evidence source & Key observation & Design implication \\
\midrule
Is there communication opportunity? & Payload replay over real DDP bucket schedules & Low-bit replay shows substantial modeled bucket-communication reduction across the evaluated schedules & Compact payloads create opportunity, but replay alone does not establish reducer feasibility. \\
Do reducer semantics matter? & Grouped timeline comparison & Hierarchical sign-count aggregation avoids the replicated worker-side counting of stock packed-sign transport & Low-bit aggregation needs reducer semantics, not only payload packing. \\
When does endpoint service dominate? & Topology and endpoint-capacity sweeps & Wins dominate the sweep space, while losses concentrate under fast networks, high fan-in, and low endpoint service rate & Reducer admission must be capacity-aware. \\
Does the guard change the route decision? & Guarded versus forced endpoint-bound replay & Guarded fallback stays near the FP32 path, whereas forced low-bit queues at the endpoint & Endpoint-capacity checks should be part of runtime route semantics. \\
Does contention change the value of payload reduction? & Contention replay with background NCCL traffic & Payload reduction remains useful under measured contention factors, but the endpoint boundary remains visible & Multi-tenant settings can increase opportunity without removing the guard. \\
Is the sign-count data path executable? & Distributed hierarchical sign-count execution & Aggregate-output checks return zero mismatches across the tested bucket executions & The reducer path is valid as an executable data path. \\
Is there component-level timing evidence? & Local-count route measurement on two nodes & Two coalesced logical-bucket shapes return zero mismatches; local-count route means are 260 us for 8 x 131K and 570 us for 8 x 524K & Component results support feasibility but are not a production speedup claim. \\
\bottomrule
\end{tabularx}
\end{table}

\subsection{Evaluation Environment And Measurement Boundary}

Table~\ref{tab:app-eval-environment} summarizes where each evidence class is measured and what is modeled. The purpose is to separate real closed-loop training evidence from reducer replay and component evidence. The paper uses the replay and sweep tools to reason about reducer admission, but accuracy and route-correctness claims come from executable PyTorch DDP runs.

\begin{table}[t]
\caption{Evaluation environment and measurement boundary.}
\label{tab:app-eval-environment}
\scriptsize
\setlength{\tabcolsep}{3pt}
\begin{tabularx}{\linewidth}{YYYY}
\toprule
Evidence class & Execution environment & Measured directly & Modeled or replayed quantity \\
\midrule
Closed-loop vision training & PyTorch DDP/NCCL on Narval A100 GPU nodes; CIFAR-100/ResNet-18 over two- and four-node deployments & Final and best top-1 accuracy, low-bit/fallback/FP32 path counts, Supervisor events, route-correctness checks & Modeled gradient-communication traffic normalized to FP32 from realized route counts \\
Transformer runtime and accuracy & PyTorch DDP/NCCL on Narval A100 GPU nodes; SST-2 with pretrained DistilBERT and BERT-base follow-ups over one- and two-node deployments & Validation accuracy, veto/readmission behavior, low-bit route share, mixed-bucket route checks, FP32 matched baselines & Modeled gradient-communication traffic normalized to matched FP32 execution \\
Autoregressive language-model runtime validation & PyTorch DDP/NCCL on Narval A100 GPU nodes; Pythia-1.4B and Pythia-6.9B on WikiText-103 raw text with full-parameter DDP over 4 nodes x 4 A100 GPUs & Training loss, validation loss/perplexity, matched FP32 anchors, route shares, route-correctness checks, hook event counts & Modeled gradient-communication traffic from the realized route log \\
Reducer payload replay & Captured DDP bucket schedules and measured collective payload curves from the evaluated workloads & Bucket sizes, bucket ordering, and baseline communication curves & Expected reducer communication time under the hierarchical sign-count model \\
Endpoint-capacity and topology sweeps & Offline sweep over node count, GPUs per node, endpoint service bandwidth, and inter-node bandwidth & Sweep inputs and guard decisions under the declared model & Reducer-admission region: whether the modeled reducer route beats the matched FP32 route \\
Contention replay & Measured background NCCL stress levels replayed against ResNet/CIFAR and Transformer/SST-2 bucket schedules & Contention factors from background all-reduce stress settings & Reducer benefit under light, medium, and heavy communication interference \\
Executable reducer checks and component microbenchmarks & Distributed sign-count and local-count measurements on two Narval nodes with four A100 GPUs per node & Aggregate-output correctness, visible traffic ratio, and local-count route timings & Not an end-to-end training-throughput model; used only as reducer-path feasibility evidence \\
\bottomrule
\end{tabularx}
\end{table}

\clearpage

\subsection{Transformer And Autoregressive Language-Model Evidence Support}

Tables~\ref{tab:app-transformer-controller-relaxation} and~\ref{tab:app-transformer-supervisor-followups} split the Transformer controller evidence into two parts. Table~\ref{tab:app-transformer-controller-relaxation} reports the controller-relaxation ladder used to move from a conservative transferred profile toward probe-guided online control. Table~\ref{tab:app-transformer-supervisor-followups} reports follow-up Supervisor checks that test whether recovery can distinguish near-threshold noise from stronger degradation.

\begin{table}[t]
\caption{Transformer controller ladder for profile-guided online control.}
\label{tab:app-transformer-controller-relaxation}
\scriptsize
\setlength{\tabcolsep}{3pt}
\begin{tabularx}{\linewidth}{YYYY}
\toprule
Controller setting & Result recorded & What the result isolates & Main-text use \\
\midrule
Fixed-profile short-gate baseline & 86.93\% accuracy, 2.34\% low-bit, 0.986x MGTR, one Supervisor trigger, no readmissions & The real-data Transformer hook path is correct, but the transferred profile is conservative on a short run & Establishes the baseline that motivates controller relaxation. \\
Fixed-profile extended-horizon baseline & 87.61\% accuracy, 1.82\% low-bit, 0.989x MGTR, one Supervisor trigger, no readmissions & Adding epochs alone does not overcome the absolute cooldown; the controller remains conservative & Rules out training-horizon length as the only explanation. \\
Run-aware cooldown and readmission & 86.70\% accuracy, 13.54\% low-bit, 0.918x MGTR, 22 Supervisor triggers, no readmissions & Capping cooldown by the remaining run horizon raises useful low-bit route share while preserving the same binding and hook path & Shows that cooldown/readmission are active control levers. \\
No-veto traffic explorer & 88.65\% accuracy, 75.00\% low-bit, 0.548x MGTR, Supervisor veto disabled & A much higher low-bit traffic opportunity exists under the same binding, but it is an oracle-style ablation rather than the default safe controller & Bounds the opportunity that guarded control should approach without adopting no-veto execution. \\
Online probe-guided controller & 85.78\% accuracy, 22.92\% low-bit, 0.862x MGTR, 42 Supervisor triggers, one readmission, learned target 0.41, auto cooldown 6 steps & Warm-up/probe evidence can be converted into an online cooldown/readmission rule without hand-entering the final low-bit target & Supports the profile-guided online-control claim for Transformer runs. \\
\bottomrule
\end{tabularx}
\end{table}

\begin{table}[t]
\caption{Transformer Supervisor follow-up checks. All rows use the same SST-2/DistilBERT runtime path and pass route-correctness checks.}
\label{tab:app-transformer-supervisor-followups}
\scriptsize
\setlength{\tabcolsep}{3pt}
\begin{tabularx}{\linewidth}{YYYY}
\toprule
Supervisor check & Result recorded & What the result isolates & Main-text use \\
\midrule
Noise-aware Supervisor recovery & 87.92\% mean accuracy, 32.51\% mean low-bit, 0.804x MGTR, 1.33 mean noise-recovery events & Near-threshold Supervisor triggers can be handled with short recovery without changing the profile or binding & Supports the smart-Supervisor follow-up in Section~7. \\
Healthy-window target growth & 87.00\% mean accuracy, 22.74\% mean low-bit, 0.863x MGTR & Growth-only adaptation remains conservative under this trigger pattern & Motivates stronger trigger classification. \\
Combined target adaptation & 88.30\% mean accuracy, 24.00\% mean low-bit, 0.855x MGTR & Adding grow/shrink logic does not recover the noise-aware residence level & Motivates phase- and noise-aware readmission. \\
\bottomrule
\end{tabularx}
\end{table}

Table~\ref{tab:app-transformer-support} complements Tables~\ref{tab:app-transformer-controller-relaxation} and~\ref{tab:app-transformer-supervisor-followups} by reporting evaluation breadth rather than additional controller ablations. Matched references are grouped with the corresponding controlled run so that the table records what each workload family contributes without repeating the controller ladder.

\begin{table}[t]
\caption{Transformer and autoregressive language-model evidence supporting bounded transfer and scale-out runtime validation.}
\label{tab:app-transformer-support}
\scriptsize
\setlength{\tabcolsep}{3pt}
\begin{tabularx}{\linewidth}{YYYY}
\toprule
Evidence group & Evaluation setting & Key result & Role in paper \\
\midrule
DistilBERT controller with matched anchors & SST-2, pretrained DistilBERT, 12 epochs, three seeds & 87.88\% mean accuracy, 29.12\% low-bit, 0.824x MGTR; matched FP32 mean accuracy is 88.34\%, and the fixed-profile reference reaches 3.39\% low-bit at 0.980x MGTR & Main accuracy-bearing Transformer result; separates controller benefit from workload and horizon. \\
Controlled deployment movement & Topology-only, dataset/step-scale, and combined two-node checks & 87.58--88.42\% mean accuracy, 24.98--27.52\% low-bit, 0.834--0.849x MGTR & Tests the same binding and controller path under bounded cluster/runtime changes. \\
Batch-side sensitivity & Batch 64 and 256 side points around batch 128 & 88.00/87.84\% mean validation accuracy and 29.49/23.31\% low-bit route share & Checks nearby batch settings without claiming arbitrary batch invariance. \\
BERT-base encoder validation with FP32 anchor & SST-2, pretrained BERT-base, eight epochs, three seeds & 90.06\% mean accuracy, 30.27\% low-bit, 0.777x MGTR; matched FP32 mean accuracy is 89.07\% & Extends the accuracy-bearing Transformer evidence to a larger encoder model. \\
Pythia-1.4B autoregressive LM route audit & WikiText-103 raw text, four nodes x four A100 GPUs & 1.41B parameters; 96 steps; loss/perplexity 2.457/11.67 versus 2.458/11.69 FP32; 30.21\% low-bit; 0.758x MGTR; zero route violations & Runtime-scale decoder-only language-model validation. \\
Pythia-6.9B autoregressive LM route audit & WikiText-103 raw text, four nodes x four A100 GPUs & 6.86B parameters; 24 steps; loss/perplexity 2.937/18.86 versus 2.939/18.90 FP32; 29.17\% low-bit; 0.743x MGTR; zero route violations & Larger autoregressive language-model runtime check. \\
\bottomrule
\end{tabularx}
\end{table}

\clearpage

\bibliographystyle{ACM-Reference-Format}
\bibliography{references}

\end{document}